\documentclass{article} % For LaTeX2e
\usepackage{iclr2022_conference,times}

\usepackage{amsmath,amsfonts,bm}

\def\eqref#1{equation~\ref{#1}}
\def\1{\bm{1}}

\def\vx{{\bm{x}}}

\def\vz{{\bm{z}}}

\def\mX{{\bm{X}}}

\def\mZ{{\bm{Z}}}

\DeclareMathAlphabet{\mathsfit}{\encodingdefault}{\sfdefault}{m}{sl}
\SetMathAlphabet{\mathsfit}{bold}{\encodingdefault}{\sfdefault}{bx}{n}

\usepackage{hyperref}
\usepackage{url}
\usepackage{graphicx}
\usepackage{booktabs}       % professional-quality tables
\usepackage{multicol,multirow,makecell}

\usepackage[normalem]{ulem}

\newcommand{\tscbase}{\textsc{Base}}
\newcommand{\tscbig}{\textsc{Large}}
\newcommand{\libri}{Librispeech}
\newcommand{\librilt}{Libri-Light}
\newcommand{\libriltsz}{LL-60k}
\newcommand{\librisz}{LS-960}
\newcommand{\libriunsz}{LS-860}
\newcommand{\teacher}{teacher}
\newcommand{\student}{student}

\title{SPIRAL: Self-supervised Perturbation-\\Invariant Representation Learning for\\ Speech Pre-Training}

\author{Wenyong Huang, Zhenhe Zhang, Yu Ting Yeung, Xin Jiang, Qun Liu \\
Huawei Noah's Ark Lab \\
\texttt{\{wenyong.huang,zhangzhenhe1,yeung.yu.ting\}@huawei.com} \\
\texttt{\{jiang.xin,qun.liu\}@huawei.com} \\
}

\iclrfinalcopy % Uncomment for camera-ready version, but NOT for submission.
\begin{document}

\maketitle

\begin{abstract}
% We introduce a new approach for speech pre-training based on a \teacher-\student\ framework, named as Self-supervised Perturbation-Invariant Representation Learning (SPIRAL). 
We introduce a new approach for speech pre-training named SPIRAL which works by learning denoising representation of perturbed data in a \teacher-\student\ framework.
% We call our approach SPIRAL for Self-supervised Perturbation-Invariant Representation Learning (SPIRAL). 
Specifically, given a speech utterance, we first feed the utterance to a \textit{\teacher} network to obtain corresponding representation. Then the same utterance is perturbed and fed to a \textit{\student} network. The \student\ network is trained to output representation resembling that of the \teacher. At the same time, the \teacher\ network is updated as moving average of \student's weights over training steps. In order to prevent representation collapse, we apply an in-utterance contrastive loss as pre-training objective and impose position randomization on the input to the \teacher. SPIRAL achieves competitive or better results compared to state-of-the-art speech pre-training method wav2vec 2.0, with significant reduction of training cost (80\% for \tscbase{} model, 65\% for \tscbig{} model).
% Most existing works pre-train and evaluate speech models on data recorded in clean condition only. 
Furthermore, we address the problem of noise-robustness that is critical to real-world speech applications. We propose multi-condition pre-training by perturbing the student's input with various types of additive noise.
% SPIRAL demonstrates its robustness to noisy speech (xxx). The code will be released after publication.
% For real-world speech applications, noise-robustness is critical.
% We further investigate multi-condition pre-training in the SPIRAL framework by perturbing the input of \student{} with various types of additive noise. 
We demonstrate that multi-condition pre-trained SPIRAL models are more robust to noisy speech (9.0\% - 13.3\% relative word error rate reduction on real noisy test data), compared to applying multi-condition training solely in the fine-tuning stage.
%The code will be released after publication.
Source code is available \footnote{https://github.com/huawei-noah/Speech-Backbones/tree/main/SPIRAL}.
\end{abstract}

%%% todo:
% refine caption of table and figure 
% table caption autoref
% SpecAug intro refine, training config
% LM
% analysis of comparison with supervised, semi-supervised
% cite BN, LayerDrop

\section{INTRODUCTION}
%\vspace{-1mm}

% labeled data are expensive
Industrial-scale automatic speech recognition (ASR) systems are usually trained with ten-thousands of hours of hand-transcribed speech data \citep{galvez2021the}.
However, labeling speech data is expensive and time-consuming, especially for languages with small speaker populations, or for specific domains (e.g., legal, financial, scientific).

% introduce self-training, pre-training methods
Recently, methods of utilizing unlabeled speech data to improve speech recognition system have achieved remarkable progress.
Amongst them, \textit{self-training} \citep{manohar15_interspeech,Kahn2020,synnaeve2020end,chen20m_interspeech,xu20b_interspeech,park2020improved, Xiao2021contrative}, also known as \textit{pseudo-labeling}, starts by training an ASR model with labeled speech data, which is referred to as teacher model. Then the teacher model, usually combined with a language model (LM), is used to produce pseudo-labels for unlabeled speech data. Finally, the labeled data and the pseudo-labeled data are combined to train a new model, which is referred to as student model. The process is repeated by taking the student model as the teacher in next iteration. Another line of work is speech \textit{pre-training} \citep{oord2019representation, Chung2020, Wang2020, wav2vec2, Liu2020}. Pre-training learns speech representation from unlabeled data in a self-supervised way. The pre-trained model is then fine-tuned on the labeled data. Self-training and pre-training are complementary as shown in recent work \citep{Xu2021,zhang2020pushing}.

% introduce SPIRAL
In this paper, we introduce a new speech pre-training method which works by learning denoising representation of perturbed data with the \teacher-\student\ framework, named as Self-supervised Perturbation-Invariant Representation Learning (SPIRAL).
Compared to state-of-the-art speech pre-training methods such as wav2vec 2.0~\citep{wav2vec2} and HuBERT~\citep{hubert}, our method allows end-to-end training with a single contrastive loss, and without relying on discrete unit discovery techniques such as vector quantization~\citep{Jegou2011,baevski2020vqwav2vec,wav2vec2} or iterative clustering process~\citep{hubert}.
%SPIRAL also allows to combine with multi-condition training~\citep{Seltzer2013,ko15_interspeech} 
We apply multi-condition training with SPIRAL~\citep{Seltzer2013,ko15_interspeech} to improve noise-robustness for the downstream speech tasks. 
% . Data augmentation techniques such as multi-condition training with additive noise and room simulation are found to improve environmental robustness of ASR \citep{ko15_interspeech,kim17_interspeech,Ko2017}.

%% motivation
SPIRAL is motivated by the observation that human tolerates speech perturbations or distortions fairly well. For example, people can communicate effectively in a noisy environment, or over a distorted telephone channel.
Therefore, we hypothesize that by learning representation invariant to perturbation, the model will learn high-level representation which can enhance speech applications.
%  will be more reliable and robust to the speech applications such speech recognition.

%% training method
To learn perturbation-invariant representation in a self-supervised way, we employ a teacher-student framework similar to \citet{Tarvainen2017}.
During pre-training, given a speech utterance, we guide the \student{} network which consumes the perturbed utterance to learn from the \teacher{} network which consumes the clean utterance.
The \student{} is trained to produce denoised representation of the perturbed utterance similar to \teacher{}'s representation of the clean utterance.
Meanwhile, the \teacher{}, which shares the same model architecture with \student{}, is updated as moving average of the \student{}'s weights over past training steps.

We apply the \textit{in-utterance} contrastive loss to avoid model collapse to trivial constant representation~\citep{Chopra2005}. As speech utterance are sequential data, there is another possible trivial solution which we call \textit{positional collapse}. Positional collapse occurs when the \student{} ``cheats'' by exploiting position correlation in \teacher{}'s representation to minimize the loss, while ignoring the content of the input utterance.
To prevent positional collapse, we propose \textit{position randomization} by adding random number of paddings on both sides of input utterance to the \teacher{}.

Large-scale speech pre-training is computationally demanding. To reduce computation cost, we adopt a gradual down-sampling strategy in SPIRAL model, which has been verified effective in speech recognition literatures with negligible performance degradation~\citep{Peddinti2018, han20_interspeech, huang2020conv}. We also speculate that aggressive down-sampling helps to remove redundancy in speech.

% We experimentally show that SPIRAL requires significant less training time to achieve comparable or better performance than the state-of-the-art pre-training methods including wav2vec 2.0.
%% results
To evaluate the effectiveness of SPIRAL, we conduct experiments on LibriSpeech and Libri-Light datasets.
By training a small convolutional classifier on the representation of a frozen SPIRAL model, we can achieve WER of 3.5\% and 6.4\% on \libri{} test-clean and test-other respectively.
SPIRAL achieves competitive or better results compared to state-of-the-art speech pre-training methods, while being much more training-efficient.
% todi: results summary here
We also demonstrate that multi-condition pre-trained SPIRAL are more robust to noisy speech with 9.0\% - 13.3\% relative word error rate (WER) reduction on real noisy test data from ChiME-3 \citep{Barker2015}, compared to the model applying multi-condition training solely in fine-tuning stage.

%%% results on librspeech, librilight
%%% results with freeze represeantons
%%% results with noise-robust training.

\section{Related work}
%\vspace{-1mm}

\textit{Mean Teacher} (MT)~\citep{Tarvainen2017} proposes using a student network to learn from a teacher network which is the moving average version of the student in the semi-supervised learning setting. The authors apply a supervised loss for labeled data and a consistency loss between teacher and student predictions for unlabeled data.
% which allows labeled data and unlabeled data to be trained together.
However, direct application of MT to self-supervised learning leads to representation collapse~\citep{Grill2020}.

\textit{Noisy student training} (NST)~\citep{xie2019self, park2020improved} is a self-training method. NST demonstrates the importance of the aggressive injection of noise into the student. Although not emphasized, no noise is injected into pseudo-labeling process of the teacher. 
We consider our work as an extension of self-training approach to the self-supervised learning regime. Instead of using the teacher to provide pseudo-labels, we utilize the teacher for pseudo-reference representation.

% todi: refine
\textit{Denoising autoencoders}~\citep{dae2008} learn to recover a clean input from a corrupted version. 
However, speech data contain redundancy which is irrelevant to some speech applications such as speech recognition. 
Previous work~\citep{baevski2019effectiveness} shows that speech pre-training by recovering masked input speech features is not effective.
In SPIRAL, we instead enforce latent representation of a corrupted input to resemble that of the corresponding clean input.

\textit{Bootstrap Your Own Latent} (BYOL)~\citep{Grill2020} is a self-supervised image representation learning method. The method is based on a teacher-student framework similar to MT. The authors refer to student network as online network and teacher network as target network. They observe that naive application of MT to self-supervised learning leads to trivial constant representation. They prevent the representation collapse by appending a predictor to the student network. The theory behind is under investigation~\citep{chen2021exploring, tian21a}.
Our method draws inspirations from BYOL and shares the similar architecture, but there are crucial differences.
Instead of learning a single global representation for an image as in BYOL, SPIRAL learns a sequence of representation for an utterance. We aim for sequence applications such as speech recognition.
In our preliminary experiments, we observe that appending a predictor to student network is not sufficient to prevent trivial constant representation for sequential representation learning.
We use in-utterance contrastive loss~\citep{wav2vec2} combined with input position randomization to successfully avoid representation collapse.
We still keep the predictor in SPIRAL, but only for the sake of performance improvement from our observation.
Another difference is that BYOL does not perform representation denoising. BYOL applies perturbation, which they call augmentation, to both the inputs of the teacher and the student. 
We demonstrate that representation denoising is crucial for speech pre-training. When perturbation is applied to the teacher's input, the effectiveness of speech pre-training degrades drastically.

% todi: cite MCT
% This property allows SPIRAL to utilize wide varieties of speech data augmentation techniques, such as multi-condition training by adding additive noise to improve noise robustness in speech recognition \citep{ko15_interspeech}.

%% wav2vec 2.0
\textit{Wav2vec 2.0}~\citep{wav2vec2} is a self-supervised speech representation learning method which belongs to the masked prediction family. Masked prediction methods are effective for text pre-training~\citep{DevlinCLT19}, but not for speech pre-training when naively applied~\citep{baevski2019effectiveness}. The reason is that speech data contains redundancy such as speaker information, pronunciation variations, which are irrelevant to the semantic meaning of the utterance.
% while text is a natural discrete modality.
% Discrete targets are preferred over continuous targets as 
To overcome this problem, wav2vec 2.0 perform masking in intermediate latent space and performs target discretization with a differentiable quantization scheme. 
% The method utilizes in-utterance contrastive loss to prevent model collapse. Empirical experiment results show that in-utterance contrastive loss with negative samples from the same utterance performs better than drawing negative samples from different utterances of the same batch.
% Target quantization contributes to performance improvement of wav2vec 2.0.
However, quantization leads to a more complex model by introducing additional hyper-parameters and an additional diversity loss.
SPIRAL does not utilize quantization, and still achieves competitive performance compared to wav2vec 2.0.
We hypothesize that aggressive down-sampling and learning by matching output representation may help to remove redundancy from the learned representation.
We leave the investigation of whether target discretization could further improve SPIRAL for future work.

\citet{liang2018learning} demonstrates that under the supervised learning setting, enforcing noise-invariant representation by penalizing difference between clean and noisy data improves ASR model accuracy. 
% The trained models are robust to mismatched noise condition.
%  The technique also improves convergence speed during model training.

\section{METHOD}
\label{method}
%\vspace{-1mm}

\subsection{Self-supervised Perturbation-Invariant Representation Learning (SPIRAL)}

\begin{figure}[th]
   \centering
   \includegraphics[trim=6.6cm 4.1cm 8.1cm 3.6cm,clip, width=0.55\linewidth]{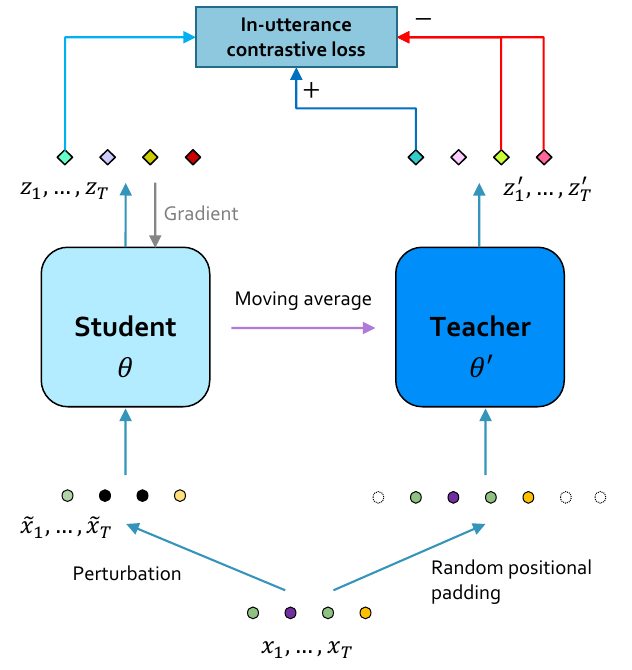}
   \caption{Illustration of SPIRAL architecture for speech pre-training.}
   \label{fig:spiral_architecture}
\end{figure}

Figure~\ref{fig:spiral_architecture} shows the diagram of SPIRAL in the pre-training stage, 
where we use two neural networks, a \textit{\student} $F_\theta$ and a \textit{\teacher} $F_{\theta'}$.
The weights of the \teacher\ $\theta'$ is the moving average of the weights of the \student\ $\theta$. At step $t$, the weights of the \teacher\ $\theta'_t$ are updated as
\begin{equation}
\theta'_t \leftarrow \alpha_t\theta'_{t-1} +  (1-\alpha_t)\theta_t,
\label{eqn:ema_update}
\end{equation}
where $\alpha_t$ determines the rate of weight updates.
% We increase the value of $\alpha_t$ from $0.995$ to $1.0$ with a cosine schedule following \citet{Grill2020} during training.
Given a speech utterance $\mX=(
\vx_1, \ldots, \vx_T)$ of length $T$, the \student\ takes a perturbed version $\tilde{\mX} = s(\mX) = (\tilde{\vx}_1, \ldots, \tilde{\vx}_T)$ as input where $s(\cdot)$ is a perturbation function. 
% One typical example of $s(\cdot)$ is the masking function commonly adopted in pre-trained language models.
The output of the student is a representation sequence $\mZ = F(\tilde{\mX};\theta) = (\vz_1, \ldots, \vz_T)$.
The \teacher\ takes the same utterance without perturbation as input and output another representation sequence $\mZ' = F(\mX;\theta') = (\vz'_1, \ldots, \vz'_T)$.
For each representation $\vz_i \in \mZ$, the \student\ is trained to match the \teacher{}'s\ representation $\vz'_i$ at the same position amongst $k$ distracting samples.
The distracting samples are randomly drawn from other positions of the same utterance in $\mZ'$, which is found to be more effective than samples drawn from an entire batch of utterances~\citep{wav2vec2}. 
The in-utterance contrastive loss is defined following \citet{sohn2016improved,wu2018unsupervised} as,
\begin{equation}
   \mathcal{L} = -\sum_{i=1}^T {\log \frac{\exp(\phi(\vz_i, \vz'_i)/\kappa)}{\sum_{j \in D_i} \exp(\phi(\vz_i, \vz'_j)/\kappa)}},
\label{eqn:contrastive_loss}
\end{equation}
%\begin{equation}
%   \mathcal{L} = -\log \frac{\exp(\phi(\vz_i, \vz'_i)/\kappa)}{\sum_{\bm{\tilde{z}'} %\sim \bm{\tilde{Z}_i}} \exp(\phi(\vz_i, \bm{\tilde{z}'})/\kappa)}
%\label{eqn:contrastive_loss}
%\end{equation}
where $\phi(\mathbf{a},\mathbf{b}) = \mathbf{a}^{T} \mathbf{b} / \|\mathbf{a}\| \|\mathbf{b}\|$ is cosine similarity, $D_i$ is the set of indices of distractors for the $i$-th position, and $\kappa$ is the temperature parameter.

However, applying in-utterance contrastive loss could cause a kind of representation collapse which we refer to as positional collapse.
Contrastive candidates are sampled based on their positions in utterances. 
When a \teacher{}'s representation $\vz'_i$ is correlated with its position $i$ (e.g., correlation introduced by positional encoding in Transformer), the \student{} could exploit this correlation to generate its representation $\vz_i$ solely based on the position index $i$, while ignoring content of the input.
% The \teacher{} and the \student{} can conspire to 'cheat' by computing and comparing a representation according to position of the candidate with respect to the utterance, instead of content of the candidate. Due to limited choice of position in the utterance of finite length, predicting the corresponding position is not a difficult task. 
In this case, 
% though a low loss value can be achievable, 
the model does not learn meaningful representation of the input content.
% When calculating the contrastive loss, we exclude the corresponding representation of the padded data.}
% todi: refine
Therefore, we prevent positional collapse by randomizing positions of teacher's representation. In particular, we add random number of padding data at both ends of the input to the teacher to randomly shift the position information for each output representation $\vz'_i$. The \student{} thereby is unable to exploit the spurious position information to minimize the contrastive loss. Note that when calculating the contrastive loss, we exclude the corresponding representation of the padded data.
% todi: pading number distrubtion

\subsection{Model architecture}
%\vspace{-1mm}

\begin{figure}[th]
%\vspace{-2mm}
   \centering
   \includegraphics[trim=0.5cm 7.5cm 0.5cm 7.2cm,clip, width=1\linewidth]{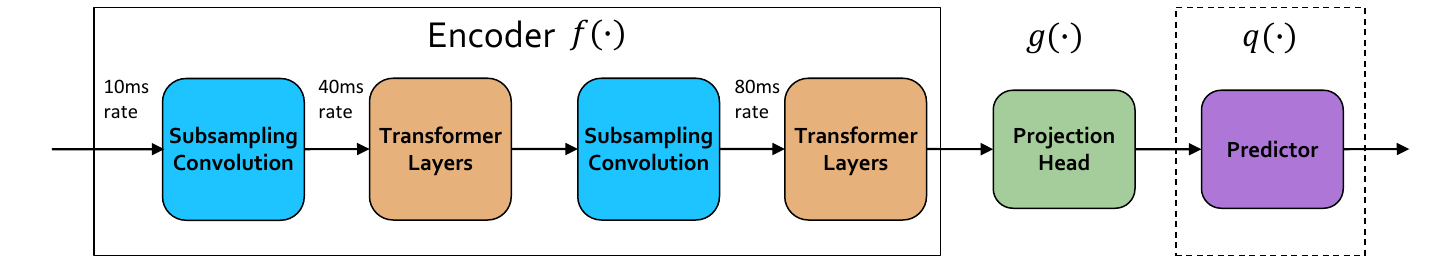}
   \caption{The architecture of the student model in SPIRAL. The frame rate of input is denoted as `10/40/80 ms'. The dashed line indicates the optional predictor which can be removed with small performance degradation. The structure of the teacher model is the same but without the predictor.}
   \label{fig:spiral_model}
\end{figure}

As illustrated in Figure \ref{fig:spiral_model}, \student\ $F_\theta$ is composed of an encoder $f(\cdot)$, a projection head $g(\cdot)$~\citep{chen2020simple} and an optional predictor $q(\cdot)$~\citep{Grill2020}, i.e., $F_\theta = (f\circ g\circ q)(\cdot;\theta)$. The \teacher\ $F_{\theta'}$ has the same structure expect that it has no predictor, $F_{\theta'} = (f\circ g)(\cdot;\theta') $.
The encoder consists of two blocks. In each block, we first apply temporal convolutions to perform down-sampling, followed by Transformer~\citep{NIPS2017_3f5ee243} with convolutional relative position encoding~\citep{wav2vec2}. Each convolution is followed by layer normalization (LN)~\citep{ba2016layer} and ReLU.
For the projection head, we apply a simple linear layer.
The predictor consists of two layers of temporal convolution and a linear layer. The convolutions are followed by batch normalization (BN)~\citep{batchnorm} and ReLU. 

During pre-training, we add computation noise to both the \student{} and the \teacher{} by applying dropout~\citep{srivastava2014dropout} and LayerDrop~\citep{fan2019reducing} in Transformer. We use the same dropout and LayerDrop rates for the \student{} and the \teacher{}.
% todo: cite BN
% We do not observe improvement with more complex projection head when there is a predictor.
% When we do not use predictor, we observe performance improvement with a temporal convolution layer with ReLU activation, followed by a linear layer for the projection head. 

\subsection{Adaptive SpecAugment}
%\vspace{-1mm}

We apply adaptive SpecAugment similar to \citet{large_scale_spec_aug} as the primary perturbation method.
Along either time or frequency dimension, we sample uniformly a certain proportion $p$ of all time-steps to be start indices and mask the subsequent consecutive $L$ time-steps.
%The number of masks $N$ is adaptive to size of each dimension.
% In SPIRAL, we apply masking on the input features, in both time and frequency domains.
The masked time-steps is filled with zeros along frequency dimension. Along time dimension, we use Gaussian noise as masking values to avoid numerical problems for LN~\citep{large_scale_spec_aug}.

\subsection{Multi-condition Pre-training}
%\vspace{-1mm}
% addtive noise
% The second type of perturbation is additive noise.
For noise-robust pre-training with SPIRAL, we perturb input of the \student{} with various types of additive noise.
% todi cite MCT 
We consider this technique as an implementation of multi-condition training (MCT)~\citep{Seltzer2013} in self-supervised setting.
Specifically, for each input utterance to the \student{}, we sample noise clips from a noise dataset, and mix the noise clips with the whole utterance by addition in time-domain. We first uniformly sample a signal-to-noise ratio (SNR) from a pre-defined range for each utterance. Then we scale the noise volume according to the required SNR. In our preliminary experiments, we found that applying additive noise alone as perturbation degrades performance. Therefore, we apply additive noise perturbation together with adaptive SpecAugment.
% We apply MCT on top of SpecAugment with 50\% probability.

\subsection{Model Fine-tuning}
%\vspace{-1mm}

After pre-training, we take the encoder from the \student{} model in SPIRAL and add a randomly initialized convolutional classifier on top of it.
The convolutional classifier is composed of two layers of convolution, followed by LN and ReLU, and a linear output projection.  The convolution filters consist of 512 channels with kernel width of 5.

We fine-tune the model with connectionist temporal classification (CTC)~\citep{Graves2006} objective for speech recognition.
We use 1024 subwords as output units. The sub-words are generated from training transcripts of LibriSpeech with SentencePiece~\citep{kudo2018sentencepiece}.
% todo: Specaug
%Aside from fined-tuning the whole model as in previous work, we also experiment with freezing the SPIRAL parameters and only fine-tuning the convolutional classifier on the top to investigate whether SPIRAL can learn high-level semantic representation of speech during pre-training. 

We further investigate SPIRAL's ability to learn high-level representation of speech during pre-training. In addition to whole-model fine-tuning, we apply frozen fine-tuning. We freeze the pre-trained parameters and only fine-tune the convolutional classifier which can only perform local classification due to limited receptive field.
% todo refine
%Since we didn't use any technique such as discrete unit discovery to remove redundancy from input speech, we want to know whether SPIRAL can learn high-level semantic representation of speech or just capturing irrelevant details of speech.
%To investigate this problem, we try another fine-tuning scheme. We freeze the pre-trained parameters and consider the SPIRAL as feature extractor and only train the output convolutional classifier, and see how this simple classifier performs.
%Note that the receptive field of the classifier is only 720ms, so all the long context modeling is done in the feature extractor.

\section{EXPERIMENTAL SETUP}
\label{experimental_setup}
%\vspace{-1mm}

\subsection{Data}
%\vspace{-1mm}

For pre-training, we use the 960-hour training data (ignoring the labels) from LibriSpeech~\citep{panayotov2015librispeech}(LS-960), or 60k-hour unlabeled audio data from Libri-Light~\citep{kahn2020libri} (LL-60K).
For Libri-Light, we segment the data using official tools with a threshold of 16s, resulting in 46.9k hours of data.
The two datasets are both derived from English audiobooks from LibriVox project\footnote{\url{https://librivox.org/}}.
For ASR fine-tuning, we apply 100-hour subset (train-clean-100) as low-resource labeled data and %960-hour data (\texttt{train-960}) 
entire LS-960 with labels
as high-resource labeled data, both from LibriSpeech.

For multi-condition training, we use the noise dataset from \citet{ms_dns}. The dataset consists of 181 hours of noise data with about 150 noise types and 70,000 clips.
We shuffle and split the noise data with a ratio of 8:1:1, which are used for training, synthesizing noisy dev-sets and synthetic noisy test-sets (results in Appendix \ref{app:synnoise}) respectively. 
SNRs of speech mixtures are set from 0 to 30 dB. 
% We also synthesize additional noisy test-sets at -10 to 0 dB as the mismatched noisy test sets.
We evaluate on real noisy data test set from CHiME-3 \citep{Barker2015}, which is comprised of speech data recorded in real noisy environments (bus, cafe, pedestrian area, and street junction). The data are recorded with a microphone array composed of multiple microphone channels located at different positions of a tablet, and a close-talking microphone. %Signal-to-noise ratios (SNR) of the data recorded from different channels are therefore different.

\subsection{Training setups}

% todo should intro mel earlier
We apply 128-dimensional log-mel filterbank extracted with 20~ms window and 10~ms stride as the input acoustic feature.
%% model setup
We experiment with \tscbase{} model and \tscbig{} model configurations as shown in Table \ref{tab:model_config}. The numbers of parameters are comparable to wav2vec 2.0 \tscbase{} and \tscbig{} models correspondingly. 
% perturbation
For SpecAugment, we set $p=0.025$ and $L=20$ for time-dimension mask, and $p=0.02$ and $L=20$ for frequency-dimension mask.

% \begin{table}[th]
%   \caption{Model configuration for SPIRAL \tscbase{} and \tscbig{}}
%   \label{tab:model_config}
%   \centering
%   \small{
%   \begin{tabular}{cc|cc}
%   \toprule
%   & & \tscbase{} & \tscbig{} \\
%   \midrule
%   \multirow{3}{*}{CNN 1}
%   & kernel width & 5, 5, 1 & 5, 5, 1 \\
%   & channel & 384, 512, 512 & 384, 512, 512 \\
%   & stride & 2, 2, 1 &  2, 2, 1 \\
%   \midrule
%         \multirow{5}{*}{Transformer 1}
%         & layer          & 2 & 4 \\
%         & embedding dim. & 512 & 512 \\
%         & inner FFN dim. & 2048 & 2048 \\
%         & layerdrop prob  & 0 & 0.05 \\
%         & attention heads & 8 & 8 \\
%   \midrule
%   \multirow{3}{*}{CNN 2}
%   & kernel width & 5, 1 & 5, 1 \\
%   & channel & 1536, 768 & 2048, 1024 \\
%   & stride & 2, 1 &  2, 1 \\
%   \midrule
%         \multirow{5}{*}{Transformer 2}
%         & layer          & 10 & 20 \\
%         & embedding dim. & 768 & 1024 \\
%         & inner FFN dim. & 3072 & 4096 \\
%         & layerdrop prob  & 0.05 & 0.05 \\
%         & attention heads & 12 & 16 \\
%   \midrule
%   \multirow{1}{*}{Proj. Head} & dim. & 256 & 512 \\
%   \midrule
%   \multirow{3}{*}{Predictor}
%   & kernel width & 5, 5, 1 & 5, 5, 1 \\
%   & channel & 256, 256, 256 & 512, 512, 512 \\
%   \midrule
%   \multicolumn{2}{c|}{Num. of Params} & 91.5M & 287M \\
%   \bottomrule
%   \end{tabular}
%   }
%   \end{table}
  
\begin{table}[t]
   \caption{Detailed configurations of the SPIRAL \tscbase{} and \tscbig{} models.}
   \label{tab:model_config}
   \centering
   \scalebox{0.9}{
   \begin{tabular}{c|ccccccc}
   \toprule
    Modules & Conv.1 & Transf.1 & Conv.2 & Transf.2 & Proj. H. & Predictor & \#Params \\
  \midrule
   \multirow{5}{*}{\shortstack{Hyper\\ -params}}
   &           & layer     &  & layer        & & &  \\
   & kernel size & emb. dim. & kernel size & emb. dim. &      & kernel size & \\
   & channel      & ffn dim. & channel       & ffn dim.  & dim. & channel & \\
   & stride       & layerdrop & stride       &  layerdrop &&&\\
   &  & attn. heads & & attn. heads &&&\\
   \midrule
   \multirow{5}{*}{\shortstack{\tscbase{}\\model}}
   &             & 2    &          & 10   &     &             &       \\
   & 5,5,1       & 512  & 5,1      & 768  &     & 5,5,1       &       \\
   & 384,512,512 & 2048 & 1536,768 & 3072 & 256 & 256,256,256 & 91.5M \\
   & 2,2,1       & 0    & 2,1      & 0.05 &     &             &       \\
   &             & 8    &          & 12   &     &             &       \\
   \midrule
   \multirow{5}{*}{\shortstack{\tscbig{}\\model}}
   &             & 4    &          & 20   &     &             &       \\
   & 5,5,1       & 512  & 5,1      & 1024 &     & 5,5,1       &       \\
   & 384,512,512 & 2048 & 2048,1024& 4096 & 512 & 512,512,512 & 287M \\
   & 2,2,1       & 0.05 & 2,1      & 0.05 &     &             &       \\
   &             & 8    &          & 16   &     &             &       \\
   \bottomrule
   \end{tabular}
   }
  \end{table}

 \begin{table}[t]
   \caption{Comparison of pre-training cost between  wav2vec 2.0 and SPIRAL.}
   \label{tab:training_cost}
   \centering
   \scalebox{0.99}{\small
   \begin{tabular}{lcccc}
   \toprule
   Model & Unlabeled data & Training steps & GPU days & Mixed precision \\
   \midrule
   Wav2vec 2.0 \tscbase{}~\scriptsize{\citep{wav2vec2}}  & \librisz{} & 500k & 102.4 & \checkmark  \\
   SPIRAL \tscbase{}  & \librisz{} & 200k &  20.8  & - \\
   \midrule
   Wav2vec 2.0 \tscbig{}~\scriptsize{\citep{wav2vec2}}  & \libriltsz{} & 1000k & 665.6 & \checkmark  \\
   SPIRAL \tscbig{}  & \libriltsz{} & 500k &  232.0 & - \\
   \bottomrule
   \end{tabular}
   }
  \end{table}
  
% training
In pre-training, we optimize with Adam~\citep{kingma2017adam} optimizer, warming up the learning rate for the first 8\% of updates to a peak of 3e-3. Then the learning rate decays to 0 with a cosine schedule. The moving average update rate $\alpha_t$ of teacher's weight also follows a cosine schedule~\citep{Grill2020}. %For \tscbase{} model, we increase the value of $\alpha_t$ from $0.995$ to $1.0$. For \tscbig{} model, the value of $\alpha_t$ from $0.990$ to $0.999$.
We increase $\alpha_t$ from $0.995$ to $1.0$ and from $0.990$ to $0.999$ for \tscbase{}  and \tscbig{} models respectively.
% \hwy{(todo: PT: lr, wd, batch size, gpu num value)}
% \hwy{(todo: FT: Mask, SNR, dropout, layer drop, crop, vocab)}
We train the \tscbase{} model with batch size of 24 per GPU for 200k steps on 16 V100 GPUs, which takes about 1.3 days.
For the \tscbig{} model, we train with batch size of 20 per GPU for 500k steps on 32 V100 GPUs, which takes about 7.25 days.
As shown in Table~\ref{tab:training_cost}, there is a significant reduction of training cost (GPU days) compared to wav2vec 2.0~\citep{wav2vec2}. SPIRAL requires 80\% and 65\% less training cost for \tscbase{} and \tscbig{} respectively.
Note that mix-precision training is not applied for SPIRAL yet. %We expect further speed-up when mix-precision training is implemented.

For fine-tuning, we optimize with Adam and a tri-state rate schedule where the learning rate is warmed up for the first
10\% of updates to 3e-5, held constant for the next 40\% and then linearly decayed to zero following \citet{wav2vec2}. We fine-tune \tscbase{} and \tscbig{} with batch size of 14 and 18 per GPU respectively on 8 GPUs for 80k steps on train-clean-100. We fine-tune \tscbig{} with batch size of 10 per GPU on 16 GPUs for 320k steps on LS-960. We apply SpecAugment for whole-model fine-tuning but not for frozen fine-tuning.
% todo: add specaug config, maybe in appendix
%For multi-condition training, 
For multi-condition pre-training and fine-tuning, 
we randomly perturb each utterance with additive noise with 50\% probability before applying SpecAugment. SNR is uniformly sampled from 0-30 dB. 

\subsection{Language Model and Decoding}

We use a word-level Transformer LM~\citep{baevski2018adaptive} trained on \libri{} LM corpus which is identical to \citet{synnaeve2020endtoend}. For low-resource ASR setting, we also evaluate SPIRAL \tscbase{} with the official LibriSpeech 4-gram LM.
We observe that models fine-tuned with subword units performs worse than models fine-tuned with character units when decoding with word-level LM.
Therefore, we apply character-based models for LM decoding, which is the same setting as wav2vec 2.0. 
The results of LM decoding with subword-based models are available in Appendix~\ref{app:lm_decoding_units_comparison}.

As output frame rate of pre-trained SPIRAL encoder is low (80ms), the output sequence may be too short for character units. To reuse the pre-trained encoder, we devise an upsampling strategy for the SPIRAL encoder output in fine-tuning stage.
We apply a 1-D convolution layer to project the original encoder output of dimension $d$ into a vector of dimension $4d$. At each time-step, we reshape the projected output vector from $(1, 4d)$ to $(4, d)$. The frame rate now becomes 20ms. 
Then we feed the upsampled outputs to convolutional classifier. 

We perform random search for decoding parameters and choose the best parameters according to performance on dev-other with beam 50.
The final test performance is measured with beam 500.
% Finally, the best parameters are applied to decoding on all the test sets with beam 500.
We use the beam search decoder of \citet{pratap2019w2l}.

\section{RESULTS}
\label{results}

\subsection{Evaluation under Low-Resource and High-Resource Labeled Data Settings}

\begin{table}[t]
 \caption{ASR results fine-tuned from low-resource train-clean-100. Language models used in decoding are listed in LM. 
 We compare SPIRAL \tscbase{} pre-trained on \librisz{} and SPIRAL \tscbig{} pre-trained on \libriltsz{} with previous methods. 
 We report WER (\%) on \libri{} dev/test sets.}
 \label{tab:low_resource}
 \centering
 \scalebox{0.98}{\small
 \begin{tabular}{lccrrrrr}
 \toprule
 \multirow{2}{*}{Model} & Unlabeled & \multirow{2}{*}{LM} & \multicolumn{2}{c}{dev} && \multicolumn{2}{c}{test} \\
 \cline{4-5}\cline{7-8}
 {} & data & {} & clean & other && clean & other \\
 \midrule
 \midrule
 \textbf{Supervised/Semi-Supervised} \\
%  \multicolumn{8}{l}{\textbf{100h labeled}}\\
 Hybrid DNN/HMM \scriptsize{\citep{L_scher_2019}} & - & 4-gram & 5.0 & 19.5 && 5.8 & 18.6 \\
%  TTS data augm.~\scriptsize{\citep{aleks2020need}} & - & LSTM & & && 4.3 & 13.5 \\
%  Discrete BERT~\scriptsize{\citep{baevski2019effectiveness}} & \librisz{} & 4-gram & 4.0 & 10.9 && 4.5 & 12.1 \\
 Iter. pseudo-labeling~\scriptsize{\citep{xu20b_interspeech}} 
%  & \libriunsz{} & 4-gram+Transf. & 4.98 & 7.97 && 5.59 & 8.95 \\
 & \libriltsz{} & 4-gram+Transf. & 3.19 & 6.14 && 3.72 & 7.11 \\
 Noisy student~\scriptsize{\citep{park2020improved}} & \libriunsz{} & LSTM & 3.9 & 8.8 && 4.2 & 8.6 \\
\midrule\midrule
   \textbf{Self-supervised} \\
 wav2vec 2.0 \tscbase{}~\scriptsize{\citep{wav2vec2}} & \librisz{} & - & 6.1 & 13.5 && 6.1 & 13.3 \\
%  && 4-gram & 2.7 & 7.9 && 3.4 & 8.0 \\
 SPIRAL \tscbase{} frozen (ours)  & \librisz{} & - & 7.9 & 12.7 && 7.6 & 13.0 \\
 SPIRAL \tscbase{} (ours)  & \librisz{} & - & 5.5 & 11.1 && 5.4 & 11.2 \\
 \midrule
 wav2vec 2.0 \tscbase{}~\scriptsize{\citep{wav2vec2}} & \librisz{} & 4-gram &  2.7 &  7.9 &&  3.4 & 8.0 \\
 SPIRAL \tscbase{} (ours)  & \librisz{} & 4-gram & 2.7 & 7.0 && 3.3 & 7.5 \\
 \midrule
 wav2vec 2.0 \tscbase{}~\scriptsize{\citep{wav2vec2}} & \librisz{} & Transf. & 2.2 & 6.3 && 2.6 & 6.3 \\
 SPIRAL \tscbase{} (ours)  & \librisz{} & Transf. & 2.3 & 5.8 && 2.7 & 6.1 \\
  \midrule
 wav2vec 2.0 \tscbig{}~\scriptsize{\citep{wav2vec2}} & \libriltsz{} & - & 3.3 & 6.5 &&  3.1 & 6.3 \\
 SPIRAL \tscbig{} frozen (ours)  & \libriltsz{} & - & 7.1 &  9.2 && 6.6 & 9.7 \\
 SPIRAL \tscbig{} (ours)  & \libriltsz{} & - &  3.3 & 5.9 && 3.3 &  6.3 \\
    \midrule
 wav2vec 2.0 \tscbig{}~\scriptsize{\citep{wav2vec2}} & \libriltsz{} & Transf. & 1.9 & 4.0 && 2.0 & 4.0 \\
 SPIRAL \tscbig{} (ours) & \libriltsz{} & Transf. & 1.9 & 3.9 && 2.2 & 4.3 \\
 \bottomrule
 \end{tabular}
 }
\end{table}

We first evaluate our method under a low-resource ASR setting in which we fine-tune the models with 100-hour LibriSpeech data (train-clean-100). 
The results are shown in Table~\ref{tab:low_resource}. 
We evaluate a \tscbase{} model pre-trained with 960-hour LibriSpeech (LS-960) and a \tscbig{} model pre-trained with Libri-Light (LL-60K).
The frozen \tscbase{} model performs well, achieving a WER of 13.0\% on test-other, which is on par with wav2vec 2.0 \tscbase{}. This suggests that SPIRAL indeed learns meaningful high-level representations in a self-supervised way.
When we fine-tune the whole \tscbase{} model, the model achieves WER of 5.4\% and 11.2\% on test-clean and test-other respectively, outperforming wav2vec 2.0 \tscbase{} with 11.5\% and 15.8\% relative WER reduction.
When decoding with Transformer LM, 
the \tscbase{} model achieves WER of 2.7\% and 6.1\% on test-clean and test-other respectively. The results are on par with wav2vec 2.0 \tscbase{}.

%\sout{However, when decoding with a Transformer LM, the gain of SPIRAL is smaller than that of wav2vec 2.0 and leads to inferior performance.
%The transformer LM is word based, but the SPIRAL models are operating in subword level. The mapping of words to subwords is not unique. This could hinder beam search performance.
%We expect that subword-level LM and more decoding parameter tuning can close the performance gap. }

The SPIRAL \tscbig{} model consists of more parameters and is pre-trained with more data. 
The model achieves WER of 2.2\% and 4.3\% on test-clean and test-other respectively. The significant improvement of \tscbig{} over \tscbase{} demonstrates the scalability of SPIRAL. 
The results of SPIRAL \tscbig{} are competitive to wav2vec 2.0 \tscbig{}.
This is encouraging, as SPIRAL \tscbig{} only takes 35\% of training cost of wav2vec 2.0 \tscbig{}.

% todi: present results
We further evaluate SPIRAL \tscbig{} pre-trained with Libri-Light (LL-60K) under a high-resource ASR setting with 960-hour LS-960 as fine-tuning data.
As shown in Table~\ref{tab:high_resource}, the \tscbig{} model achieves WER of 1.8\% and 3.5\% on test-clean and test-other respectively, which are on par with the wav2vec 2.0 \tscbig{} model.
%\sout{We again observe the performance improvement of SPIRAL with LM is smaller than wav2vec 2.0. 
%We except to close the performance gap with the mentioned LM decoding refinement.}
We note that the supervised models and the noisy student model \citep{park2020improved} in Table~\ref{tab:high_resource} are autoregressive models.
Our models are fine-tuned with CTC objective which is non-autoregressive and generally inferior to autoregressive models.
We use CTC objective for its simplicity and comparability to previous speech pre-training methods. 

We consider SPIRAL as a preferred alternative to wav2vec 2.0 given that SPIRAL only requires 20\%$-$35\% computation cost of wav2vec 2.0. We expect further efficiency improvement when we implement mix-precision training for SPIRAL.

\begin{table}[t]
   \caption{%
   ASR results fine-tuned from high-resource LS-960. Language models used in decoding are listed in LM. 
   We compare SPIRAL \tscbig{} pre-trained on \librilt{}  (\libriltsz{}) with previous methods. We report WER (\%) on \libri{} dev/test sets.}
   \label{tab:high_resource}
   \centering
   \scalebox{0.98}{\small
   \begin{tabular}{lccrrrrr}
   \toprule
   \multirow{2}{*}{Model} & Unlabeled & \multirow{2}{*}{LM} & \multicolumn{2}{c}{dev} && \multicolumn{2}{c}{test} \\
   \cline{4-5}\cline{7-8}
   {} & data & {} & clean & other && clean & other \\
   \midrule\midrule
   \textbf{Supervised} \\
%   CTC Transf~\scriptsize{\citep{synnaeve2020end}} & - & CLM+Transf. & 2.20 & 4.94 && 2.47 & 5.45 \\
%   S2S Transf.~\scriptsize{\citep{synnaeve2020end}} & - & CLM+Transf. & 2.10 & 4.79 && 2.33 & 5.17 \\
%   Transf. Transducer~\scriptsize{\citep{zhang2020transformer}} & - & Transf. & - & - && 2.0 & 4.6 \\
   ContextNet~\scriptsize{\citep{han2020contextnet}} & - & LSTM & 1.9 & 3.9 && 1.9 & 4.1 \\
   Conformer~\scriptsize{\citep{gulati2020conformer}} & - & LSTM & 2.1 & 4.3 && 1.9 & 3.9 \\
   \midrule\midrule
   \textbf{Semi-supervised} \\
   CTC Transf. + PL~\scriptsize{\citep{synnaeve2020end}} & \libriltsz{} & CLM+Transf. &  2.10 & 4.79 && 2.33 & 4.54 \\
   S2S Transf. + PL~\scriptsize{\citep{synnaeve2020end}} & \libriltsz{} & CLM+Transf. &  2.00 & 3.65 && 2.09 & 4.11 \\
   Iter. pseudo-labeling~\scriptsize{\cite{xu20b_interspeech}} & \libriltsz{} & 4-gram+Transf. & 1.85 & 3.26 && 2.10 & 4.01 \\
   Noisy student~\scriptsize{\citep{park2020improved}} & \libriltsz{} & LSTM & 1.6 & 3.4 && 1.7 & 3.4 \\
   \midrule\midrule
   \textbf{Self-supervised} \\
   wav2vec 2.0 \tscbig{}~\scriptsize{\citep{wav2vec2}} & \libriltsz{} & - & 2.1 & 4.5 && 2.2 & 4.5 \\
   SPIRAL \tscbig{} frozen (ours) & \libriltsz{} & - & 4.0 & 6.2 &&  3.5 &  6.4 \\
   SPIRAL \tscbig{} (ours) & \libriltsz{} & - & 2.1 & 4.3 && 2.2 &  4.6 \\
   \midrule
    wav2vec 2.0 \tscbig{}~\scriptsize{\citep{wav2vec2}} & \libriltsz{} & Transf. & 1.6 & 3.0 && 1.8 & 3.3 \\
    SPIRAL \tscbig{} (ours) & \libriltsz{} & Transf. & 1.5 & 3.1 && 1.8 &  3.5 \\
   \bottomrule
   \end{tabular}
   }
   \end{table}

\subsection{Noise-robust pre-training}

\begin{table}[t]
   \caption{Evaluation on noise-robustness of the models. We use wav2vec 2.0 \tscbase{} released by the authors as the baseline. The SPIRAL \tscbase{} models are pre-trained with \librisz{} and fine-tuned with train-clean-100. We report WER (\%) on \libri{} and CHiME-3 real data test sets.}
   \label{tab:noisy_test}
   \centering

  \scalebox{0.98}{\small
   \begin{tabular}{lccrrrrrrrrrrr}
   \toprule
   \multirow{2}{*}{\tscbase{} model} & Pre-train & Fine-tune & \multicolumn{2}{c}{Librispeech} && \multicolumn{4}{c}{CHiME-3} \\
   \cline{4-5}\cline{7-10}
   {} & w/ MCT & w/ MCT & clean & other && ch0 & ch5 & ch1 & ch2 \\
   \midrule
   wav2vec 2.0 & -         & -      & 6.1 & 13.3 && 23.2 & 56.1 & 68.3 & 98.1 \\
   SPIRAL & -         & -          & 5.4 & 11.2 &&  24.1 & 52.1 & 58.9 & 92.6 \\
   SPIRAL & -         & \checkmark & 5.7 & 11.4 && 20.8 & 35.5 & 41.1  & 76.4\\
   SPIRAL & \checkmark & -         & 5.7 & 11.5 && 20.8 & 33.6 & 38.5 & 74.0 \\
   SPIRAL & \checkmark & \checkmark &  5.9 & 11.4 && 20.0 & 31.1 &  35.6 & 69.5 \\
   \bottomrule
   \end{tabular}

  }
\end{table}
% \small{
% \begin{tabular}{cccrrrrrrrrrrrrr}
%   \toprule
%   \multirow{2}{*}{\tscbase{} model} & Pre-train & Fine-tune & \multicolumn{2}{c}{SNR 0-30 dB} && \multicolumn{2}{c}{SNR -10-0 dB} && \multicolumn{2}{c}{Original} \\
%   \cline{4-5}\cline{7-8}\cline{10-11}
%   {} & w/ MCT & w/ MCT & clean & other && clean & other && clean & other \\
%   \midrule
%   wav2vec 2.0 & -         & -          & 14.4 & 27.4 && 67.7 & 78.3 && 6.1 & 13.3 \\
%   SPIRAL & -         & -          & 13.1 & 24.3 && 61.2 & 71.4 && 6.0 & 11.9 \\
%   SPIRAL & -         & \checkmark & 8.64 & 17.7 && 35.5 & 51.4 && 6.4 & 12.0 \\
%   SPIRAL & \checkmark & -          & 8.0 & 16.1 && 28.2 & 44.0 && 6.3 & 12.0 \\
%   SPIRAL & \checkmark & \checkmark & 8.0 & 15.8 && 26.1 & 41.4 &&  6.6 & 12.1 \\
%   \bottomrule
%   \end{tabular}
% }

To evaluate noise-robustness of the pre-trained models, we compare the effects of applying multi-condition training (MCT) in pre-training or fine-tuning stages of SPIRAL.
The results are shown in Table~\ref{tab:noisy_test}.
The vanilla SPIRAL \tscbase{} model and wav2vec 2.0 \tscbase{} model deteriorate with significantly higher WER on noisy test data. 
% Applying MCT in pre-training or fine-tuning stage all improve fine-tune stage is more noise-robust, has lower WER on Noisy tests. 

% We further observe that applying additive noise perturbation at pre-training stage is more effective than in fine-tuning stage, achieving higher speech recognition accuracy.
%When using noise-Perturbation in fine-tuning in addition to per-training, we found no significant improvement. This shows the noise-robustness can be transferred efficiently to downstream applications.
% When performing MCT for both pre-training and fine-tuning, the further improvement is rather limited 

% The results suggest that noise-robustness achieved in SPIRAL pre-training is effectively transferred to the downstream ASR tasks. 
% Nevertheless, fine-tuning with MCT is still helpful when we test on real noisy test data. We observe more significant improvement at the mismatched SNR range (-10-0 dB), with 7.4\% and 6\% for relative WER reduction for test-clean and test-other respectively.
On real noisy test speech data in CHiME-3 for different microphone channels (ch), SPIRAL with multi-condition pre-training significantly improves speech recognition performance.  
%The results further show that multi-condition pre-training significantly improves the performance for real noisy test speech data in CHiME 3 for different microphone channels (ch) except ch 0, which is a close-talking microphone with the highest SNR.
Compared to the model applying MCT solely in fine-tuning, applying MCT both in pre-training and fine-tuning achieves 12.4\%, 13.3\% and 9.0\% relative WER reduction for ch 1, 5 and 2 respectively. There is smaller performance improvement of 3.8\% relative WER reduction for ch 0, which is a close-talking microphone with the highest SNR. We note that ch 2 faces backwards to the speaker. SNR of the recordings from ch 2 is the lowest, leading to high WER. 
We note that other pre-training methods including wav2vec 2.0 may benefit from multi-condition training, which are worth for further investigation.
%We further observe that for noisy test data, applying MCT solely on pre-training stage tends to achieve lower WER than applying MCT solely on fine-tuning stage.
% We only test for a subset (0,1,2,5) of microphone channels. The other microphones (3,4,6) are on similar relative positions to the speakers on the tablet. We expect the similar performance trends as mic 1 and mic 5.

\subsection{Ablations}
\label{sec:Ablations}

\subsubsection{Input Perturbation and computation noise of teacher}

%\begin{table}[t]
%   \caption{Evaluating effects of input perturbation with SpecAugment and computation noise on teacher. We list the mask portion and mask length as $p$, $L$ for time and frequency masks. The first row is our default setting in which we do not apply SpecAugment to teacher's input. We use SPIRAL \tscbase{} for testing.}
%   \label{tab:perturb_test}
%   \centering
%   % \setlength\tabcolsep{3.3pt}
%%   \small{
%   \begin{tabular}{cccr}
%   \toprule
%   Time mask & Frequency mask & Computation noise & dev other \\
%   \midrule
%   -       &   -       & \checkmark & 11.7 \\
%   -       &   -       & - & 13.9 \\
%   0.025, 20    &  0.02, 20 &  \checkmark  & 53.1 \\
%   0.0125, 20  &   0.01, 20 & \checkmark  & 47.0 \\
%   0.025, 10 &    0.02, 10  & \checkmark   & 42.8 \\
%   \bottomrule
%   \end{tabular}
%%   }
%  \end{table} 

SPIRAL learns denoised representation of perturbed data. 
By default, we only apply perturbation to the input of the \student{}. 
An alternative method is to perturb both inputs of the \teacher{} and the \student{}, and optimize consistency between their representations~\citep{Grill2020, chen2021exploring}.
We conduct experiments to evaluate the effects of perturbing the input and adding computation noise (dropout and LayerDrop) to the \teacher{}.
The results are shown in Table~\ref{tab:perturb_test} in Appendix \ref{app:ablation}. The results suggest that applying SpecAugment to \teacher{}'s input degrades performance significantly.
Performance degradation decreases but is still significant with lower ratio and width of the masks.
This supports the necessity of representation denoising, and our view of SPIRAL as an extension of self-training in which \teacher{} network are fed with clean input. 
% We should apply noise to the input of the student, but not the input of the teacher.
The results also support applying computation noise to teacher during pre-training. There is a 15.9\% relative WER reduction with computation noise. This may be linked to \citet{dropbayes}.

\subsubsection{Effects of predictor and projection head}
%\begin{table}[ht]
%   \caption{Effects of predictor and projection head.  We use SPIRAL \tscbase{} for testing. We report WER (\%) on the \libri{} dev other set.}
%   \label{tab:predcitor_ablations}
%   \centering
%   % \setlength\tabcolsep{3.3pt}
%%   \small{
%   \begin{tabular}{lr}
%   \toprule
%   Architecture & dev other \\
%%   \cline{2-3}
%%   & clean & other  \\
%   \midrule
%   baseline                   & 11.7 \\
%   + predictor use LN    & 12.3 \\
%   + conv proj. head  & 12.3 \\
%   -- predictor                         & 14.6 \\
%   \enskip\enskip + conv proj. head     & 12.2 \\
%%   \enskip\enskip + temperature 0.4     & - & 13.3 \\
%%   \enskip\enskip\enskip + learning rate 2e-3     & - & 13.1 \\
%   \bottomrule
%   \end{tabular}
%%   }
%  \end{table}

We do ablation studies to understand the role of predictor and projection head in SPIRAL.
The results are shown in Table~\ref{tab:predcitor_ablations} in Appendix \ref{app:ablation}.
When removing the predictor from the \student{}, we observe performance degradation, but representation collapse does not happen.
In the architectures relying on predictor to prevent collapse~\citep{Grill2020, chen2021exploring},
applying batch normalization (BN) in the predictor is essential. While in SPIRAL, we observe that BN in the predictor can be replaced by layer normalization (LN) with a small performance degradation.
%can also improve performance, though not as much as those using BN.
When the predictor is removed, we observe performance improvement by applying a convolutional projection head.
The convolutional projection head is composed of a temporal convolution layer with LN and ReLU, and a linear layer. 
But applying convolutional projection head to the model with a predictor, there is no further performance improvement. This suggests that convolutional projection head and predictor play a similar role in SPIRAL, and they are not complementary.

\section{CONCLUSION}
\label{conclusion}
%\vspace{-2mm}
We presented SPIRAL, a new approach to speech pre-training by learning denoising representation of perturbed data with a teacher-student framework.
SPIRAL can learn high-level speech representation in self-supervised way. Training a small convolutional classifier on frozen representation of SPIRAL achieves WER of 3.5\% and 6.4\% on \libri{} test-clean and test-other respectively.
We show that SPIRAL achieves competitive or better results compared to state-of-the-art speech pre-training methods, with significant reduction of training cost. We investigate multi-condition pre-training and demonstrates that multi-condition pre-training is more effective than solely applying multi-condition training in the fine-tuning stage. We presume SPIRAL as a general pre-training method, which can 
apply to other modalities such as images and text. We leave it for future work.
% We expect performance gains by switching to a seq2seq architecture.

\bibliography{iclr2022_conference}
\bibliographystyle{iclr2022_conference}

\newpage
\appendix
\section{Appendix}

\subsection{Output units comparison for Language Model Decoding}
\label{app:lm_decoding_units_comparison}

We evaluate decoding performance of different combinations of SPIRAL output units and Transformer LM. The subword-level LM is trained on the same Librispeech LM corpus and shares the same 1024 subwords as the SPIRAL model fine-tuned with subword units.

\begin{table}[h]
 \caption{ASR results fine-tuned from low-resource train-clean-100 and high-resource train-960. The model units and language models for decoding are listed in fine-tuning units and LM respectively. We compare SPIRAL \tscbase{} pre-trained on \librisz{} and SPIRAL \tscbig{} pre-trained on \libriltsz{} with previous methods.
 We report WER (\%) on \libri{} dev/test sets.}
 \label{tab:lm_decoding_units}
 \centering
 \scalebox{1}{%\small
 \begin{tabular}{lcccrrrrr}
 \toprule
 \multirow{2}{*}{Model} & Unlabeled & Fine-tuning & \multirow{2}{*}{LM} & \multicolumn{2}{c}{dev} && \multicolumn{2}{c}{test} \\
 \cline{5-7}\cline{8-9}
 {} & data & units & {} & clean & other && clean & other \\
 \midrule
 \midrule
   \textbf{Low-resource} \\
 SPIRAL \tscbase{}   & \librisz{} & subword & - & 5.5 & 11.1 && 5.4 & 11.2 \\
 SPIRAL \tscbase{}   & \librisz{} & char & - & 5.3 & 11.0 && 5.4 & 11.1 \\
  SPIRAL \tscbase{}   & \librisz{} & subword & word & 2.9 & 6.8 && 3.2 & 7.2 \\
  SPIRAL \tscbase{}   & \librisz{} & subword & subword & 2.7 & 6.3 && 2.9 & 6.7 \\
 SPIRAL \tscbase{}   & \librisz{} & char & word & 2.3 & 5.8 && 2.7 & 6.1 \\
  \midrule
 SPIRAL \tscbig{}   & \libriltsz{} & subword & - &  3.3 & 5.9 && 3.3 &  6.3 \\
  SPIRAL \tscbig{}   & \libriltsz{} & char & - &  3.5 & 6.3 && 3.5 & 6.7 \\
 SPIRAL \tscbig{}  & \libriltsz{} & subword & word & 2.4 & 4.5 && 2.5 & 4.8 \\
 SPIRAL \tscbig{}  & \libriltsz{} & subword & subword & 2.3 & 4.5 && 2.4 & 4.8 \\
 SPIRAL \tscbig{}  & \libriltsz{} & char & word & 1.9 & 3.9 && 2.2 & 4.3 \\
 \midrule
 \midrule
 \textbf{High-resource} \\
   SPIRAL \tscbig{} & \libriltsz{} & subword & - & 2.1 & 4.3 && 2.2 &  4.6 \\
   SPIRAL \tscbig{} & \libriltsz{} & char & - &  2.2 & 4.5 && 2.3 &  4.7 \\
   SPIRAL \tscbig{}  & \libriltsz{} & subword & word & 1.7 & 3.5 && 1.9 & 3.7 \\
   SPIRAL \tscbig{}  & \libriltsz{} & subword & subword & 1.6 & 3.3 && 1.7 & 3.5 \\
    SPIRAL \tscbig{}  & \libriltsz{} & char & word & 1.5 & 3.1 && 1.8 &  3.5 \\
 \bottomrule
 \end{tabular}
 }
\end{table}

\subsection{Performance on synthetic noisy dataset}
\label{app:synnoise}
On the synthetic noisy dataset (NS-Librispeech) with matched SNR range (0-30 dB) of the training data, 
SPIRAL pre-trained and fine-tuned with MCT is more effective than applying MCT solely in fine-tuning. We observe 5.3\% and 9.1\% relative WER reduction on synthetic noisy test-clean and test-other sets respectively.

\begin{table}[h]
   \caption{Evaluation on noise-robustness of the models. We use wav2vec 2.0 \tscbase{} released by the authors as the baseline. The SPIRAL \tscbase{} models are pre-trained with \librisz{} and fine-tuned with train-clean-100. We report WER (\%) on \libri{} test sets and synthetic noisy \libri{} test sets at 0 - 30 dB (NS-Librispeech).}
   \label{tab:synsetic_noisy_test}
   \centering

  {
   \begin{tabular}{lccrrrrrrrrrr}
   \toprule
   \multirow{2}{*}{\tscbase{} model} & Pre-train & Fine-tune & \multicolumn{2}{c}{Librispeech} && \multicolumn{2}{c}{NS-Librispeech}\\
   \cline{4-5}\cline{7-8}
   {} & w/ MCT & w/ MCT & clean & other && clean & other \\
   \midrule
   wav2vec 2.0 & -         & -      & 6.1 & 13.3 &&  14.4 & 27.4  \\
   SPIRAL & -         & -          & 5.4 & 11.2 && 12.2 & 23.3  \\
   SPIRAL & -         & \checkmark & 5.7 & 11.4 && 7.6 & 16.5 \\
   SPIRAL & \checkmark & -         & 5.7 & 11.5 && 7.4 & 15.8 \\
   SPIRAL & \checkmark & \checkmark &  5.9 & 11.4 && 7.2 & 15.0 \\
   \bottomrule
   \end{tabular}
  
  }
\end{table}

% \subsubsection{Evaluates frozen representations and effects of temperature}

% \begin{table}[ht]
%   \caption{Evaluates frozen representations and effects of temperature on contrastive loss.  We use SPIRAL \tscbase{} for testing, which is pre-trained with the audio of \libri{} (\librisz{}) and fine-tuned with \texttt{train-clean-100}. We report WER (\%) on the \libri{} dev-other set.}
%   \label{tab:frozen_test}
%   \centering
%   % \setlength\tabcolsep{3.3pt}
%   \begin{tabular}{crr}
%   \toprule
%   \multirow{2}{*}{Temperature} & \multicolumn{2}{c}{dev other} \\
%   \cline{2-3}
%   & frozen & fine-tune \\
%   \midrule
%   0.5   & 13.3 & 11.9 \\
%   0.4   & 13.1 & 11.5 \\
%   0.3   & 12.8 & 11.7 \\
%   0.2   & 12.7 & 11.4 \\
%   0.1   & 14.3 & 12.1 \\
%   \bottomrule
%   \end{tabular}
%   \end{table}

% Recent work~\citep{wang2021understanding, wang2020hypersphere} shows temperature in Equation \ref{eqn:contrastive_loss} is related to a key property of contrastive representation learning, \textit{uniformity}. Uniformity means how well the features are uniformly distributed. Smaller temperature leads to better uniformity, but may cause the contrastive loss not tolerant to semantically similar samples.
% So we evaluates models pre-trained with different temperatures.
% The results are shown in Table~\ref{tab:frozen_test}.
% As the results show, our model performs reasonalbe in temerature range $0.1~0.5$. The performance degrades a little when the temperature is too low (0.1).

\subsection{Results of ablation studies}
\label{app:ablation}
Here are the results of ablation studies discussed in Section \ref{sec:Ablations}. 
\begin{table}[h]
   \caption{Ablation studies of input perturbation with SpecAugment and computation noise on teacher. We list the mask ratio and mask length as $p$, $L$ for time and frequency masks. The first row is the default setting of SPIRAL. We apply SPIRAL \tscbase{} fine-tuned with train-clean-100, and report WER (\%) on the \libri{} dev-other set.}
   \label{tab:perturb_test}
\centering
   \scalebox{1}{%\small
   \begin{tabular}{cccr}
   \toprule
   Time mask & Frequency mask & Computation noise & dev other \\
   \midrule
   -       &   -       & \checkmark & 11.1 \\
   -       &   -       & - & 13.2 \\
   0.025, 20    &  0.02, 20 &  \checkmark  & 47.9 \\
   0.0125, 20  &   0.01, 20 & \checkmark  & 42.8 \\
   0.025, 10 &    0.02, 10  & \checkmark   & 39.4 \\
   \bottomrule
   \end{tabular}
   }
\end{table}

\begin{table}[h]
   \caption{Ablation studies of predictor and projection head in SPIRAL.  We apply SPIRAL \tscbase{} fine-tuned with train-clean-100, and report WER (\%) on the \libri{} dev-other set.
   }
   \label{tab:predcitor_ablations}
\centering
   \scalebox{1}{%\small
   \begin{tabular}{lr}
   \toprule
   Architecture & dev other \\
%   \cline{2-3}
%   & clean & other  \\
   \midrule
   SPIRAL \tscbase{}     & 11.1 \\
   + predictor use LN    & 11.6 \\
   + conv proj. head  & 11.5 \\
   -- predictor                         & 13.7 \\
   \enskip\enskip + conv proj. head     & 12.1 \\
   \bottomrule
   \end{tabular}
   }
\end{table}

\end{document}